\documentclass[aps,twocolumn,nofootinbib,showpacs,prd]{revtex4}
\usepackage{graphicx}
\usepackage{amsfonts}
\usepackage{amssymb}
\usepackage{amsbsy}
\usepackage{amsmath}
\usepackage{latexsym}
\usepackage{natbib}
\usepackage{bm}
\usepackage{color}
\usepackage{float}

\begin{document}
\title{Acceleration from a Phase of Entropic Balance}
\author{Soumya Chakrabarti\footnote{soumya.chakrabarti@vit.ac.in}}
\affiliation{School of Advanced Sciences, Vellore Institute of Technology, \\ 
Tiruvalam Rd, Katpadi, Vellore, Tamil Nadu 632014 \\
India}

\date{\today}

\begin{abstract}
We discuss the notion of generating a cosmic inflation without any big bang singularity. It has been proved recently by Good and Linder \cite{linder} that such an expansion of the universe can be driven by quantum fluctuations embedded in vacuum. The rate of expansion is guided by a cosmological sum rule defined through the Schwarzian derivative. We explore the thermodynamic roots of Schwarzian and connect it with the surface gravity associated with an apparent horizon. In General Relativity the cosmological sum rule can be enforced only if the early universe is a Milne vacuum. We show that this restriction can be removed by considering an entropic source term in the Einstein-Hilbert action.   
\end{abstract}

\maketitle

Recently it has been proposed \cite{linder} that the cosmic expansion can be driven simply by a modified vacuum structure of the early universe. In this approach the inflation is generated from a Milne vacuum plus some quantum fluctuations. There is no requirement for a Big Bang singularity and an effective expansion is driven by a Schwarzian derivative, which has primordial connection with M\"obius and Lorentz symmetries \cite{schwarzian}, artefacts of conformal field theory. In a scenario of cosmic expansion from \textit{`nothing'} \cite{linder}, the Schwarzian is used to define a cosmological sum rule and to constrain the expansion. Given that there is no clear motivation behind the universe beginning from a hot, dense, singular state, such a proposition is indeed well-motivated. However, in light of this notion, one must re-assess the popular cosmological questions, such as the graceful exit problem \cite{guth} or the genesis of a so-called dark energy term during late-times \cite{de}. In this article we point out an important correlation between the Schwarzian and the thermodynamic variables of the expanding universe enclosed by an apparent horizon. In cosmology, the homogeneous and isotropic universe can be imagined as a thermodynamic system enclosed by an apparent horizon. It can be proved that transitions of the universe, such as from a decelerated into an accelerated phase, are nothing but second order phase transition(s) (divergence of specific heat capacity) \cite{tanima, sc}. We connect the Schwarzian of the system with the surface gravity associated with the enclosing apparent horizon. We also reproduce the condition of phase transition during a cosmic expansion using the Schwarzian. It has been proved \cite{linder} that a cosmological sum rule can only be implemented if the early universe $(t \rightarrow 0)$ is described by a vacuum solution, i.e., the Milne metric. We argue that one can go beyond this requirement by introducing an entropic source term in the Einstein-Hilbert action. \\

We consider a spatially homogeneous Friedmann-Robertson-Walker (FRW) metric $ds^2 = -dt^2 + a^2(t)\,\left[dr^2 + r^2(d\theta^2 + \sin^2\theta d\phi^2)\,\right]$. The cosmological scale factor $a(t)$ is a solution of the FRW field equations for a suitably chosen energy-momentum tensor. A growth in $a(t)$ indicates a cosmic expansion and the fractional rate of this growth is quantified using a few kinematic quantities 
\begin{equation} 
H(a) = \frac{\dot a}{a} ~~,~~ q(a) = -\frac{a\ddot a}{\dot a^2} ~~,~~ j(a) = \frac{a^2\dddot a}{\dot a^3}. 
\end{equation} 
These are popularly known as the Hubble, deceleration and the jerk parameter and they denote the first, second and the third order dimensionless growth rate of the scale factor. In this context a Schwarzian is defined as
\begin{equation}\label{eq:szdef}  
\lbrace a,t \rbrace = \frac{\dddot{a}}{\dot{a}} - \frac{3}{2} \left(\frac{\ddot{a}}{\dot{a}}\right)^2 = H^2\,\left(j-\frac{3}{2}q^2\right).  
\end{equation}
The Schwarzian derivative does not depend on the spatial curvature of the background metric tensor, i.e., Eq. (\ref{eq:szdef}) remains the same for an FRW metric with $k \neq 0$. It is closely related to the symmetry group $SL(2,R)$ \cite{schwarzian}. Imposition of a Schwarzian simply means that the system is invariant under some kind of physical transformations, such as a boost or rotation. For more discussions on this we refer to the article of Kitaev \cite{kitaev}. It finds application in the description of boundary terms of quantum black holes in Jackiw-Teitelboim gravity \cite{mertens1, mertens2}. In a system exhibiting maximal chaos, Schwarzian derivatives are used to quantify conformal symmetry breaking, for instance, in the low-energy sector of a Sachdev-Ye-Kitaev (SYK) model \cite{maldacena}. Other notable applications can be found in the description of M\"obius symmetry breaking in cosmology, characterization of stress energy tensor in analog models of black hole evaporation and entanglement entropy of black hole horizons \cite{eva, entangle}. In standard cosmology, the Schwarzian is used to enforce an integral constraint on the growth of the scale factor, much like the sum rule introduced by Good, Linder and Wilczek \cite{good} for unitary solutions. This constraint provides two informations about the cosmic expansion (i) it must begin from an empty space and (ii) it must exhibit acceleration at some epoch(s). The cosmological Schwarzian is defined as \cite{linder}
\begin{eqnarray}\label{chi} 
&& \chi = \frac{1}{2} ln (\dot{a}), \\&&
\lbrace a,t \rbrace = 2\,\left[\ddot{\chi} - \dot{\chi}^2\right] = 2\,e^{\chi}\,\frac{d}{dt}\left(e^{-\chi}\dot{\chi}\right). 
\end{eqnarray} 
Since $e^{\chi} > 0$ for all $\chi$, one defines the integral sum rule using the total derivative term as
\begin{equation}\label{sumrule}
\int_{0}^{+\infty} \dot{a}^{-1/2} \lbrace a,t \rbrace dt = 0. 
\end{equation} 

The LHS of the above equation produces  
\begin{equation}  
\int_{0}^{+\infty} \dot{a}^{-1/2}\; \lbrace a,t \rbrace dt = \left(\dot{a}^{-1/2} \frac{\ddot{a}}{\dot{a}}\right) \Bigg|_{t=0}^{t=+\infty}\, \label{eq:integ}  
\end{equation} 

upon integration. If the scale factor behaves as a power law ($a\sim t^n$) when $t\to+\infty$, then $\left(\dot{a}^{-1/2} \frac{\ddot{a}}{\dot{a}}\right) \rightarrow n^{-1/2}(n-1)t^{-(1+n)/2}$. For positive values of $n$ this will always go to zero. However, in the limit $t \rightarrow 0$ it will vanish if and only if $\ddot a = 0$. This implies a coasting universe as described by the Milne metric ($a \propto t$). 
\begin{eqnarray}\label{eq:FR1milne}
&&3\frac{\dot{a}}{a}^2 + \frac{3k}{a^2} = 8\pi G \rho, \\&&
\frac{\ddot a}{a} = -\frac{4\pi G}{3}(\rho + 3p).\label{eq:FR2milne}
\end{eqnarray}
One can easily check this by writing the FRW equation and the Raychaudhuri equation as in Eq. (\ref{eq:FR1milne}) and Eq. (\ref{eq:FR2milne}) and solve it for $\rho = p = 0$; $k < 0$. The indication is quite interesting : if the early universe originates not from a hot, dense \textit{Big Bang}, but rather as a quantum fluctuation around empty spacetime, then the cosmological sum rule remains valid. During the expansion the Schwarzian must also have a transition from positive into negative in order to be consistent with the sum rule and this ensures a transition into late-time acceleration. The proposition is generic as it relies only on the vanishing of boundary terms borne out of the integral sum rule. \\ 

We point out that to implement a cosmological sum rule consistently, one does not necessarily need an empty spacetime in the early universe. In fact, the sum rule can be extended to any other space-time metric with a generic energy-momentum tensor, provided we make a suitable modification to general relativity (GR). Imagine having an energy-momentum distribution that modifies the $G^{1}_{1}$ component of the field equations but keeps the $G^{0}_{0}$ part intact, much like an effective bulk viscosity. Then the RHS of the Raychaudhuri Eq. (\ref{eq:FR2milne}) is modified. As a consequence, one can claim that in an early universe limit, i.e., $t \rightarrow 0$, the two components of the energy-momentum distribution (the GR component and the modification) simply cancel each other out to give $\ddot a = 0$ (which is the key requirement to implement the sum rule). We construct such a modification with a generally covariant framework that unifies GR and non-equilibrium thermodynamics \cite{sc, garcia}. The Lagrangian for this system depends on generalized coordinates and the entropy $S$
\begin{equation}
\label{eq:var_prin}
\delta \int_{t_1}^{t_2} L(q, \dot{q}, S) dt = 0\,.
\end{equation}
Variation of the lagrangian leads to 
\begin{equation}
\frac{\partial L}{\partial S} (q, \dot{q}, S) \delta S = \left<F(q, \dot{q}, S), \delta q \right>,
\end{equation}
where $\left< \cdot , \cdot \right>$ is a scalar product term. The term on the RHS resembles an entropic force which modifies the usual Euler-Lagrange equations and brings in an associated phenomenological constraint \cite{garcia, garcia1, thermo}
\begin{equation}
\frac{d}{dt} \frac{\partial L}{\partial \dot{q}} - \frac{\partial L}{\partial q} = F (q, \dot{q}, S) ~,~ \frac{\partial L}{\partial S} \dot{S} = \left<F (q, \dot{q}, S), \dot{q} \right> \,.
\end{equation}
This constraint ensures that the gravitational system under consideration is thermodynamically closed, however, the formalism can also be extended for an open system \cite{thermo}. The temperature of the system is introduced as $\frac{\partial L}{\partial S} = - T$. Assuming the entropy function to be homogeneous we define the modified Einstein-Hilbert action as
\begin{equation}
\label{eq:grm}
\frac{1}{2\kappa} \int d^4x \sqrt{-g} R + \int d^4x \mathcal{L}_{m}(g_{\mu \nu}, S)\,.
\end{equation}

An extremisation of this action leads to
\begin{eqnarray}\nonumber
&& \int d^4x \left[ \left(\frac{1}{2\kappa} \frac{\delta (\sqrt{-g}R)}{\delta g^{\mu \nu}} + \frac{ \delta \mathcal{L}_m}{\delta g^{\mu \nu}}\right)\delta g^{\mu \nu} + \frac{\partial \mathcal{L}_m}{\partial S} \delta S \right] = 0, \\&&
\mathcal{L}_m \equiv \sqrt{-g} L_{m} ~~,~~ \frac{\partial L_{m}}{\partial S} \delta S = \frac{1}{2} F_{\mu \nu} \delta g^{\mu \nu}.
\end{eqnarray}

$F_{\mu \nu} = \int d^3x \sqrt{-g} f_{\mu \nu}$ is a tensor density term associated with the entropic correction. The modified field equations derived from this action are
\begin{eqnarray}\label{eq:GREA}
&& G_{\mu\nu} = \kappa \left( T_{\mu \nu} -  f_{\mu \nu} \right), \\&&
f_{\mu\nu} = \zeta\,D_\lambda u^\lambda \,(g_{\mu\nu}+u_\mu u_\nu) = \zeta\,\Theta\,h_{\mu\nu}.
\end{eqnarray}
The entropic correction $f_{\mu\nu}$ in the modified field Eq. (\ref{eq:GREA}) resembles an effective bulk viscosity, where $\zeta$ is the coefficient of bulk viscosity which we can express in terms of $\Theta$, the expansion scalar and $V$, the comoving volume. For a $k = 0$ FRW spacetime this coefficient becomes 
\begin{equation}
\zeta = \frac{T}{\Theta}\frac{dS}{dV} \equiv T\dot S/(9H^2a^3).
\end{equation}
Due to this construction the effective energy-momentum tensor behaves like an imperfect fluid 
\begin{equation}
{\cal T}^{\mu\nu} = p\,g^{\mu\nu} + (\rho + p)u^\mu u^\nu - \zeta\,\Theta\,h^{\mu\nu}.
\end{equation}
From the Bianchi identities it can be checked that the entropy-modified energy-momentum tensor is not covariant divergence free. We can write the modified field equations as

\begin{eqnarray}\label{eq:FR1}
&&3\frac{\dot{a}^2}{a^2} = 8\pi G \rho, \\&&
\frac{\ddot a}{a} = -\frac{4\pi G}{3}(\rho + 3p) + \frac{4\pi G}{3}\frac{T\dot S}{a^3H}.\label{eq:FR2}
\end{eqnarray}

It must be noted that any such entropic modification of GR can simply be realized by calculating the equation of continuity $\dot\rho + 3H(\rho+p) = \frac{T\dot S}{a^3}$, which is nothing but the modified second law of thermodynamics ($T dS = d(\rho a^3) + p\,d(a^3) = 0$). \\

The cosmological sum rule, i.e.,
\begin{equation}  
\int_{0}^{+\infty} \dot{a}^{-1/2}\; \lbrace a,t \rbrace dt = \left(\dot{a}^{-1/2} \frac{\ddot{a}}{\dot{a}}\right) \Bigg|_{t=0}^{t=+\infty} = 0,
\end{equation}
is valid in the presence of an entropic source term if $\dot{a}^{-1/2} \frac{\ddot{a}}{\dot{a}}$ vanishes in the limit $t \rightarrow 0$ as well as $t \rightarrow \infty$. We can still assume that the scale factor to behave as a power law when $t \rightarrow \infty$ and find that $\left(\dot{a}^{-1/2} \frac{\ddot{a}}{\dot{a}}\right) \rightarrow 0$. However, in early universe, i.e., $t \to 0$, $\frac{\ddot{a}}{\dot{a}}$ can vanish iff the RHS of Eq. (\ref{eq:FR2}) is zero. Therefore, in presence of an entropic source term the requirement to implement a cosmological sum rule becomes
\begin{equation}\label{condi1}
\lim_{t \to 0} \frac{T}{a^2} \frac{dS}{da} = (\rho + 3p).
\end{equation}

Eq. (\ref{condi1}) provides a modified requirement. At the outset, the condition to implement a cosmological sum rule remains the same : the universe should accommodate quantum fluctuation in an effective vacuum during the early inflation. However, this vacuum should not necessarily mean that the universe is devoid of matter. Rather it means that the contribution to the effective acceleration from standard matter is in an exact equilibrium with the entropic force term, much like Newton's third law. One can imagine the expansion of universe as a non-equilibrium phenomenon between two fixed points, both described by an effective vacuum. In principle, Eq. (\ref{condi1}) can be satisfied by any generic energy momentum tensor provided the entropy function is chosen accordingly. For instance, if the strong energy condition for the matter distribution is satisfied ($\rho + 3p > 0$), we must have a growth in entropy as a function of scale factor in the early universe. As the universe expands, due to the $\frac{T}{a^2}$ factor the growth of entropy is gradually subdued. Categorically, this evolution resembles the concept of an Entropic balance. An Entropic balance demands that within an isolated system the effective change in total entropy due to a spontaneous process can either be positive or zero; but not negative. The formalism to define the temperature $T$ must have a mention in this context. We go with the assumption that the expanding universe is enclosed by an apparent horizon. We treat this horizon as an evolving null surface, in order to account for the growth in degrees of freedom at the causal boundary. The temperature of the horizon can be defined using the Hayward-Kodama formalism \cite{hayward1, hayward2, hayward3}. This choice is more suitable for dynamic horizons compared to the more widely used Hawking temperature formalism, which is typically associated with static horizons \cite{hawking}. Other advantages of this choice have been discussed in recent literature, particularly in studies focusing on the thermodynamic stability analysis during cosmic expansion \cite{stab}. First we rewrite the FRW metric as
\begin{equation}\label{2eq3}
ds^{2} = h_{ab}dx^{a}dx^{b} + \tilde{r}^{2}d\Omega^{2},
\end{equation}
where $\tilde{r} = a(t)r$, $x^{0} = t$, $x^{1} = r$ and $h_{ab} = {\rm diag}(-1, a^{2}/1-kr^{2})$. There is a formation of apparent horizon when the vector $\nabla \tilde{r}$ is null, i.e., $h^{ab}\partial_{a}\tilde{r}\partial_{b}\tilde{r} = 0$. For the FRW metric a radius corresponding to the formation of apparent horizon can be derived as
\begin{equation}
\label{2eq4}
\tilde{r}_{A} = 1/\sqrt{H^{2}+k/a^{2}}.
\end{equation}
In the Hayward-Kodama formalism, surface gravity $\kappa$ on the area radius of two-sphere is defined using the Kodama vector $K^a \equiv \epsilon^{ab} \nabla_b R$, which obeys the equation
\begin{equation}\label{HK surgrav}
\frac{1}{2}g^{ab}K^c(\nabla_c K_a-\nabla_a K_c)= \kappa K^b.
\end{equation}
The surface gravity $\kappa$ is written as
\begin{equation}\label{2eg5}
\kappa = \frac{1}{2\sqrt{-h}}\partial_{a}(\sqrt{-h}h^{ab}\partial_{b}\tilde{r}),
\end{equation}
where $\epsilon^{ab}$ is the volume form of the induced two-metric $h^{ab}$. For an FRW metric the surface gravity at apparent horizon can be derived as
\begin{equation}\label{2eq6}
\kappa = -\frac{1}{\tilde{r}_{A}}(1 - \frac{\dot{\tilde{r}}_{A}}{2H\tilde{r}_{A}}),
\end{equation}
where the overhead dot is a cosmic time-derivative. If the geometry is spatially flat, this can be simplified further to assign the following \textit{Hayward-Kodama} temperature to the apparent horizon \cite{haykod}
\begin{equation}\label{temp}
T = \frac{\mid{\kappa\mid}}{2\pi} = \frac{2H^2+\dot{H}}{4\pi H} = \frac{H}{4\pi}\left(1-q\right).
\end{equation}
We also write the the first order time derivative of the associated temperature as 
\begin{equation}\label{tempdot}
\dot{T} =-\frac{1}{4\pi H^2}\left(q^2 + q + 1 - j \right).
\end{equation}
Using Eqs. (\ref{temp}) and (\ref{tempdot}) the Schwarzian derivative for the system can be written as
\begin{equation}\label{modischw}
\lbrace a,t \rbrace = H^2\Big[4\pi H^2 \dot{T} - \frac{1}{2}\Big(1-4\pi \frac{T}{H}\Big)^2 + \Big(1-4\pi \frac{T}{H}\Big) + 1 \Big].  
\end{equation}
Finally, using Eq. (\ref{modischw}) in Eqs. (\ref{sumrule}) and (\ref{eq:integ}) we derive that the integral sum rule demands that
\begin{equation}
\left[\left(\frac{H}{a}\right)^{\frac{1}{2}} \left(\frac{\kappa}{2H}-1\right)\right]^{\infty}_{0} = 0.
\end{equation}
This identity means that the surface gravity must be proportional to Hubble function in the limits, $t \to 0$ and $t \to \infty$ if we must implement the cosmological sum rule. Using Eq. (\ref{temp}), it is straightforward to solve the $\left(\frac{\kappa}{2H}-1\right) = 0$ equation and find a scale-factor capable of describing this behavior as a function of cosmic time. We also note the connection between a Schwarzian and a thermodynamic phase transition. We first write the horizon entropy (proportional to surface area) and its time derivative as $S_h = 2\pi A$, $\dot{S_h}= -16 \pi^2 \frac{\dot{H}}{H^3}$. $A$ is the area enclosed by the apparent horizon. We define $S_{\mbox{in}}$ and $U$ as the entropy and total internal energy of the enclosed fluid. Using the first law of thermodynamics $TdS_{\mbox{in}} = dU + pdV$ we generate a constraint for the adiabatic case. Here $V$ is the volume of the fluid enclosed by the apparent horizon, i.e., a sphere of radius $Y = \frac{1}{H}$, i.e., $V = \frac{4}{3}\pi \frac{1}{H^3}$. We calculate the first order change in entropy as $\dot{S}_{\mbox{in}} = \frac{1}{T_h} \left[(\rho+p)\dot{V} + \dot{\rho}V\right]$. Using this, we proceed with a thermodynamic stability analysis by maximizing the entropy. It involves the derivation of principle minors of a Hessian matrix of entropy \cite{karter}. The mathematical requirements for this maximization are written as
\begin{eqnarray} \nonumber
&& (i)\frac{\partial^2{S_{\mbox{in}}}}{\partial U^2} \equiv = -\frac{1}{T^2C_V} \leq 0, \label{condithermo}\\&&\nonumber
(ii)\frac{\partial^2{S_{\mbox{in}}}}{\partial U^2}\frac{\partial^2{S_{\mbox{in}}}}{\partial V^2}- \left(\frac{\partial^2{S_{\mbox{in}}}}{\partial U\partial V}\right)^2 \equiv \frac{1}{C_VT^3V\beta_T} \geq 0, \label{condi2}
\end{eqnarray}

where we have defined the specific heat capacities as
\begin{equation}\label{cv}
C_V = T\left(\frac{\partial S_{\mbox{in}}}{\partial T}\right)_V ;~ C_P = T\left(\frac{\partial S_{\mbox{in}}}{\partial T}\right)_P.
\end{equation}

We note that any divergence of the heat capacities depend on the differential change in the temperature associated with the system, in this case, the Hayward-Kodama temperature. This divergence can only be realized when
\begin{equation}\label{condiii}
\frac{\partial T}{\partial t}dT = 0 \Rightarrow \ddot{H}H + 2H^{2}\dot{H} - \dot{H}^2 = 0.
\end{equation}
Using the dimensionless parameters deceleration and jerk, we rewrite this as a kinematic condition
\begin{equation}\label{dimlesscondi}
q^2 + q + 1 - j = 0.
\end{equation}
It is remarkable to note that exactly the same condition can also be found from the Schwarzian 
\begin{eqnarray}\nonumber
&& H^2 \Big(j-\frac{3}{2}q^2\Big) = H^2\Big[4\pi H^2 \dot{T} - \frac{1}{2}\Big(1-4\pi \frac{T}{H}\Big)^2 \\&&
+ \Big(1-4\pi \frac{T}{H}\Big) + 1 \Big],  
\end{eqnarray}
by putting $\dot{T} = 0$. This gives us a generic condition for second order phase transition(s) during cosmic expansion : the time rate of change of the horizon temperature must vanish. This condition is effectively a quadratic relation between the kinematic parameters describing fractional growth of the scale factor, as in Eq. (\ref{dimlesscondi}). If we substitute $j = 1$ for a $\Lambda$CDM model, we can find from Eq. (\ref{dimlesscondi}) that the phase transition is only possible for either $q = 0$ or $q = -1$. The point $q = 0$ clearly denotes a transition point between deceleration and acceleration. We point out towards one more interesting observation taking a $\Lambda$CDM model as an example. First, we rewrite the Schwarzian Eq. (\ref{eq:szdef}) as  
\begin{equation}
H^2 \left(j-\frac{3}{2}q^2\right) = \epsilon.
\end{equation}
We recall that to have a consistent description of cosmic acceleration, apart from the vanishing boundary terms of the integral sum rule, the Schwarzian must go through a change in signature \cite{linder}. In that context, $\epsilon$ can be thought of as a parameter, which can be assigned positive or negative values depending on the phase of expansion the universe is in. If we assign the the present value of deceleration parameter to be $q \sim -2/3$, it is easy to find that the present epoch satisfies
\begin{equation}
j = \frac{2}{3} + \frac{\epsilon}{H^2}.
\end{equation}
However, if $\epsilon$ is negative (as it should during the epoch of late-time acceleration \cite{linder}) the jerk parameter can never be $1$ (which denotes a $\Lambda$ CDM model). This points out to either of the following two scenarios.
\begin{enumerate}
\item {If the Schwarzian for the present universe is assumed to be positive, a $\Lambda$CDM can be a viable solution for which the kinematic parameters are asigned a value of the order $q \sim 0.67$, $j \sim 1$. Nevertheless, $\epsilon$ will have to evolve into negative values in future to obey the integral sum rule.}
\item {If the Schwarzian has already moved into a negative quadrant during the present accelerated expansion, the $\Lambda$CDM can not be a viable solution for the same. To keep the deceleration parameter value close to the observed value $q \sim 2/3$, jerk parameter must have a value much different from $1$.}
\end{enumerate}  

We conclude the article with a note that the integral sum rule can also be related to an identity of Riemannian geometry. However, this is not unexpected, since the requirement of vanishing boundary term or $\frac{\ddot{a}}{a}$, is directly connected to the Raychaudhuri equation for an FRW space-time. One may recall that the Raychaudhuri equation governs the expansion scalar $\Theta$ as \cite{rc}
\begin{equation}
\frac{d\Theta}{d\tau} + \frac{1}{3}\Theta^{2} + \sigma^2 - \omega^2 + R_{\alpha\beta}u^\alpha u^\beta = 0.
\label{rc_eqn}
\end{equation}
$\sigma^2 = \sigma^{\alpha\beta}\sigma_{\alpha\beta}$ where $\sigma_{\alpha\beta}$ is the shear tensor. Similarly, $\omega^2 = \omega^{\alpha\beta}\omega_{\alpha\beta}$ where $\omega_{\alpha\beta}$ is the rotation tensor. The Ricci tensor is written as $R_{\alpha\beta}$. For an FRW metric $\sigma^2 = \omega^2 = 0$ and the first two terms on the LHS of Eq. (\ref{rc_eqn}) simply gives $\frac{\ddot{a}}{a}$. Therefore a requirement of vanishing boundary term implies that 
\begin{equation}
\lim_{t \to 0} R_{\alpha\beta}u^\alpha u^\beta \to 0 ~~;~~ \lim_{t \to \infty} R_{\alpha\beta}u^\alpha u^\beta \to 0.
\end{equation}
This is nothing but the convergence condition for a family of geodesics in Riemannian geometry. In GR this condition is more popularly studied as the Null energy condition. One might ask if this simple correspondence can be seen for a space-time metric other than the FRW metric or for a physical system other than homogeneous and isotropic expansion. In a way, imposing the cosmological sum rule has provided an escape from the cosmological singularity at $t \to 0$. It is therefore a natural curiosity to ask if the sum rule can be reformulated in the case of a gravitational collapse of a dying star, which usually leads to the formation of a space-time singularity. At the outset there seems to be an issue with the function $\chi$ since $\chi = \frac{1}{2} ln (\dot{a})$, which remains viable as long as $\dot{a} > 0$, i.e., only for expanding metrics. For a collapsing metric $\dot{a} < 0$ and with this in mind, we propose a revised sum rule
\begin{eqnarray}\label{chi_coll} 
&& \chi = \frac{1}{2} ln \left(\frac{\dot{a}}{\alpha}\right), \\&&
\lbrace a,t \rbrace = 2\,\left[\ddot{\chi} - \dot{\chi}^2\right] = 2\,e^{\chi}\,\frac{d}{dt}\left(e^{-\chi}\dot{\chi}\right). 
\end{eqnarray} 

$\alpha$ gives the time rate of change of the radius of two-sphere, hence it should be negative. With this, we can define the integral sum rule for all $\chi$ as
\begin{equation}\label{sumrule_coll}
\int_{t=t_i}^{t=t_f} \left(\frac{\dot{a}}{\alpha}\right)^{-\frac{1}{2}} \lbrace a,t \rbrace dt = \left\lbrace \left(\frac{\dot{a}}{\alpha}\right)^{-\frac{1}{2}} \frac{\ddot{a}}{\dot{a}}\right\rbrace \Bigg|_{t_i}^{t_f}\,= 0. 
\end{equation} 
$t_i$ and $t_f$ denote the values of time coordinate when the collapse begins and ends, respectively. If a spacetime singularity is supposed to form at $t \to t_f$, the rate of collapse should ideally approach infinity. Clearly, in this limit $\frac{\ddot{a}}{\dot{a}} \to 0$. However, any example of collapse and bounce ($\ddot{a} > 0$, $\dot{a} = 0$) produces $\frac{\ddot{a}}{\dot{a}} \to infty$ at a finite value of $t$, resulting in a general breakdown of the integral sum rule. Any gravitational collapse producing a stable state of dynamical equilibrium also leads to the same. A general analysis of the sum rule for a spherical collapsing distribution not restricted to spatial homogeneity or pressure isotropy may provide some more clarity regarding these questions and this topic will be addressed in a follow-up article.

\section*{Acknowledgement}

The author bids acknowledgement to the IUCAA for providing facilities under the visiting associateship program and to the Vellore Institute of Technology for the financial support through its Seed Grant (No. SG20230027), $2023$.


\begin{thebibliography}{99}

\bibitem{linder} M. R. R. Good and E. V. Linder, arXiv : 2503.02380 [gr-qc]. 

\bibitem{schwarzian} G. W. Gibbons, arXiv : 1403.5431 [hep-th] ; A. Kitaev, arXiv : 1711.08169 [hep-th].

\bibitem{guth} A. H. Guth, and S. Y. Pi, Phys. Rev. Lett., {\bf 49}, 1110 (1982) ; Phys. Rev. D, {\bf 32}, 1899 (1985).

\bibitem{de} A. G. Riess et al., Astron. J. {\bf 116}, 1009 (1998), Astron. J. {\bf 117}, 707 (1999) ; S. Perlmutter et al., Astrophys. J. {\bf 517}, 565 (1999) ; J. L. Tonry et al., Astrophys. J. {\bf 594}, 1 (2003) ; B. J. Barris et al., Astrophys. J. {\bf 602}, 571 (2004) ; M. Hicken et al., Astrophys. J. {\bf 700}, 1097 (2009).

\bibitem{tanima} T. Duary, N. Banerjee and A. Dasgupta, Eur. Phys. J. C., {\bf 83}, 815 (2023).

\bibitem{sc} S. Chakrabarti, Fortschr. Phys. {\bf 72}, 11 https://doi.org/10.1002/prop.202400063 (2024).

\bibitem{kitaev} A. Kitaev, arXiv : 1711.08169 [hep-th] (2017).

\bibitem{mertens1} T. G. Mertens and G. J. Turiaci, Living Rev. Rel. {\bf 26}, 4 (2023). 

\bibitem{mertens2} T. G. Mertens, JHEP {\bf 05}, 036 (2018).

\bibitem{maldacena} J. Maldacena and D. Stanford, Phys. Rev. D. {\bf 94}, 106002 (2016).

\bibitem{eva} B. S. DeWitt, Phys. Rept. {\bf 19}, 295 (1975) ; S. A. Fulling and P. C. W. Davies, Proc. R. Soc. Lond. A {\bf 348}, 393 (1976).

\bibitem{entangle} M. R. R. Good and E. V. Linder, Class. Quant. Grav. {\bf 39}, 105003 (2022).

\bibitem{good} M. R. R. Good, E. V. Linder and F. Wilczek, Phys. Rev. D {\bf 101}, 025012 (2020).

\bibitem{garcia} L. Espinosa-Portales, J. Garcia-Bellido, Phys. Dark Univ., {\bf 34}, 100893 (2021).

\bibitem{garcia1} R. Arjona, L. Espinosa-Portales, J. Garcia-Bellido and S. Nesseris, Phys. Dark Univ., 36, 101029 (2022).

\bibitem{thermo} F. Gay-Balmaz, H. Yoshimura, J. Geom. Phys. {\bf 111}, 194 (2017).

\bibitem{hayward1} S. A. Hayward, Class. Quant. Gravit., {\bf 15}, 3147 (1998).

\bibitem{hayward2} S. A. Hayward, R. Di Criscienzo, M. Nadalini, L. Vanzo and S. Zerbini, Class. Quant. Grav. {\bf 26}, 062001 (2009).

\bibitem{hayward3} R. Di Criscienzo, S. A. Hayward, M. Nadalini, L. Vanzo and S. Zerbini, Class. Quant. Grav. {\bf 27}, 015006 (2010).

\bibitem{hawking} S. W. Hawking, Commun. Math. Phys. {\bf 43}, 199 (1975).

\bibitem{stab} A. Paranjape, S. Sarkar and T. Padmanabhan, Phys. Rev. D. {\bf 74}, 104015 (2006) ; R. G. Cai and L. M. Cao, Phys. Rev. D. {\bf 75}, 064008 (2007) ; R. G. Cai and N. Ohta, Phys. Rev. D {\bf 81}, 084061 (2010); R. D’Agostino, Phys. Rev. D., {\bf 99}, 103524 (2019).

\bibitem{haykod} R. G. Cai and S. P. Kim, J. High Energy Phys. {\bf 2005}, 050 (2005).

\bibitem{karter} A. H. Carter, Am. J. Phys. {\bf 68}, 1158 (2002).

\bibitem{rc} A. Raychaudhuri. Phys. Rev., {\bf 98}(4) : 1123, (1955).









\end{thebibliography}
\end{document}